\documentclass[11pt,a4paper]{article}
\pdfoutput=1

\usepackage{jcappub}
\usepackage{mathrsfs}
\usepackage{bbm}
\usepackage{bm}
\usepackage{dsfont}
\usepackage{tensor}
\usepackage{xspace}
\usepackage{aeguill}
\usepackage{wasysym}
\usepackage{microtype}
\usepackage{empheq}
\usepackage{ifthen}
\usepackage{amsmath}
\usepackage{subcaption}
\usepackage{wrapfig}

\usepackage[table]{xcolor}
\usepackage{multirow}
\newcommand{\cst}{\mathrm{cst}}

\newcommand\define{\equiv}

\newcommand\vect[1]{\boldsymbol{#1}}
\newcommand{\mat}[1]{\boldsymbol{#1}}

\newcommand\transpose[1]{#1^{\rm T}}

\newcommand\e[1]{_{\text{#1}}}

\newcommand\U[1]{\:\mathrm{#1}}

\newcommand{\order}{\mathcal{O}}

\newcommand{\dd}{\mathrm{d}}

\newcommand{\ddf}[3][]{\frac{\dd^{#1} #2}{\dd {#3}^{#1}}}

\renewcommand\lim[2]{\underset{ #1 \rightarrow #2 }{ \mathrm{lim} } \,}

\newcommand{\delimiters}[4][]{
\ifthenelse{ \equal{#1}{1} }{  #2 #3 #4  }
					{ \ifthenelse{\equal{#1}{2}}{ \big#2 #3 \big#4 }
						{ \ifthenelse{\equal{#1}{3}}{ \Big#2 #3 \Big#4 }
							{ \ifthenelse{\equal{#1}{4}}{ \bigg#2 #3 \bigg#4 }
								{ \ifthenelse{\equal{#1}{5}}{ \Bigg#2 #3 \Bigg#4 }
									{ \left#2 #3 \right#4 }
								}
							}
						}
					}
													}

\newcommand{\pa}[2][]{\delimiters[#1]{(}{#2}{)}}
\newcommand{\pac}[2][]{\delimiters[#1]{[}{#2}{]}}

\newcommand{\ev}[2][]{\delimiters[#1]{\langle}{#2}{\rangle}}

\newcommand{\Hc}{\mathcal{H}}

\newlength{\boxtitlelength}
\newlength{\halfrulelength}
\newcommand{\boxtitle}[1]{\footnotesize\bf{\:#1\:}}

\definecolor{blue4}{RGB}{0,0,143}
\definecolor{red4}{RGB}{143,0,0}
\definecolor{orange}{RGB}{255,128,0}
\definecolor{darkcyan}{RGB}{0,128,128}
\definecolor{olive}{RGB}{0,128,0}
\definecolor{purple}{RGB}{128,0,128}
\definecolor{cyan2}{RGB}{0,255,255}
\definecolor{fushia}{RGB}{255,0,255}
\definecolor{mygray}{gray}{0.5}
\definecolor{lightgray}{gray}{0.85}

\newcommand{\kernel}{W}

\title{How does the cosmic large-scale structure bias the Hubble diagram?}

\author[a,b]{Pierre Fleury,}
\author[a,c]{Chris Clarkson,}
\author[b,d]{Roy Maartens.}

\affiliation[a]{Department of Mathematics \& Applied Mathematics, University of Cape Town,\\
Cape Town 7701, South Africa}
\affiliation[b]{Department of Physics \& Astronomy, University of the Western Cape,\\
Cape Town 7535, South Africa}
\affiliation[c]{School of Physics \& Astronomy, Queen Mary University of London,\\
327 Mile End Road, London E1 4NS, United Kingdom}
\affiliation[d]{Institute of Cosmology \& Gravitation, University of Portsmouth,\\
Portsmouth PO1 3FX, United Kingdom}

\emailAdd{pierre.fleury@uct.ac.za}
\emailAdd{chris.clarkson@qmul.ac.uk}
\emailAdd{roy.maartens@gmail.com}

\abstract{The Hubble diagram is one of the cornerstones of observational cosmology. It is usually analysed assuming that, on average, the underlying relation between magnitude and redshift matches the prediction of a Friedmann-Lema\^itre-Robertson-Walker model. However, the inhomogeneity of the Universe generically biases these observables, mainly due to peculiar velocities and gravitational lensing, in a way that depends on the notion of average used in theoretical calculations. In this article, we carefully derive the notion of average which corresponds to the observation of the Hubble diagram. We then calculate its bias at second-order in cosmological perturbations, and estimate the consequences on the inference of cosmological parameters, for various current and future surveys. We find that this bias deeply affects direct estimations of the evolution of the dark-energy equation of state. However, errors in the standard inference of cosmological parameters remain smaller than observational uncertainties, even though they reach percent level on some parameters; they reduce to sub-percent level if an optimal distance indicator is used.}


\date{\today}

\begin{document}

\maketitle
\flushbottom

\section{Introduction}
\label{sec:introduction}

The birth of the observational side of cosmology can be considered to date from Hubble's discovery of the recession of galaxies~\cite{1929PNAS...15..168H}. The Hubble diagram, showing that nearby galaxies recede with a velocity proportional to their distance, was indeed the first observational evidence of the concept of cosmic expansion, proposed independently by Friedmann~\cite{1922ZPhy...10..377F} and Lema\^itre~\cite{1927ASSB...47...49L}. Later on, with the astrophysical discovery of type Ia supernovae, and as technology allowed us to probe the cosmos to much larger distances, this same Hubble diagram, which represents the relation that exists in our Universe between luminosity distance~$d\e{L}$ and redshift~$z$, led to the discovery that cosmic expansion is accelerating~\cite{1998AJ....116.1009R,1999ApJ...517..565P}. Today's Hubble diagrams still use type Ia supernovae (SNIa)~\cite{2014A&A...568A..22B}, but also since very recently quasars (QSOs)~\cite{2015ApJ...815...33R} and gamma ray bursts~\cite{2016arXiv161000854D}, and in the future gravitational wave (GW) emitters, the so-called standard sirens~\cite{1986Natur.323..310S,2005ApJ...629...15H}.

Although instruments and data analysis made great progress in the construction of the observational Hubble diagram, the theoretical framework in which it is analysed is remarkably simple: it relies on the strictly homogeneous and isotropic cosmological model of Friedmann, Lema\^itre, Robertson, and Walker (FLRW). More precisely, it means that the relation between observed distances and redshift is the one predicted by the FLRW model. This actually represents two main assumptions: (i) the large-scale expansion dynamics of our Universe is correctly described by this model; (ii) on average, the effect of matter on light propagation is as if light beams were propagating though a purely homogeneous fluid. In this article, we assume that (i) is valid, even though it is still matter of debate---the so-called backreaction issue~\cite{2011arXiv1109.2314C,2012ARNPS..62...57B,2013arXiv1311.3787W,2014CQGra..31w4003G,2015arXiv150507800B,2015arXiv150606452G,2015PhRvL.114e1302A,2016CQGra..33l5027G,2016arXiv160903724F}. Note however that the \emph{local} departure from homogeneity has recently be shown~\cite{2016arXiv160904081E} to be potentially responsible for the problematic discrepancy between local~\cite{2016ApJ...826...56R} and global~\cite{2016A&A...594A..13P} measurements of the Hubble constant~$H_0$.

Let us focus on assumption~(ii) which concerns light propagation in our Universe. The confrontation of the optical properties of an inhomogeneous and a homogeneous Universe raises two kinds of issues. The first one is the validity of the fluid approximation: do the optical properties of a lumpy Universe match the ones of a fluid-filled Universe? This question was originally addressed in the 60s-70s by several authors~\cite{1964SvA.....8...13Z,DashevskiiZeldovich1965,1966SvA.....9..671D,1965AZh....42..863D,1966RSPSA.294..195B,1967ApJ...150..737G,1969ApJ...155...89K,DR72,1973ApJ...180L..31D,1973PhDT........17D,1974ApJ...189..167D,1975ApJ...196..671R,1967ApJ...147...61G,1976ApJ...208L...1W} and reviewed recently in refs.~\cite{Clarkson:2011br,2016MNRAS.455.4518K}. The same issue has also motivated several series of works in the past decades~\cite{Holz:1997ic, Futamase:1989hba, Clifton:2009jw,2009JCAP...10..026C,Clarkson:2011br,Bruneton:2012ru,2012PhRvD..85b3502C,2013PhRvD..87l3526F,2013PhRvL.111i1302F,2014JCAP...06..054F,2016arXiv161109275B}, until the recent proposition of a new stochastic formalism~\cite{2015JCAP...11..022F} that would be more realistic than the standard perturbation theory on such scales. In short, the conclusions are that the distance-redshift relation can be biased up to $10\%$ at redshift~$z=1$ in a Universe made of opaque clumps. This bias is dramatically reduced for transparent clumps.
The second aspect of light propagation in the inhomogeneous Universe consists in the effect of the cosmic large-scale structure. Unlike the lumpiness issue, this kind of inhomogeneity is well-described by the relativistic perturbation theory~\cite{PeterUzan}, although alternatives based on exact solutions of the Einstein equation have been exploited as well, such as Swiss-cheese models~\cite{1945RvMP...17..120E,1946RvMP...18..148E} with Lema\^itre-Tolman-Bondi~\cite{Marra:2007pm,2008PhRvD..77b3003M,Brouzakis:2007zi,2007JCAP...02..013B,Biswas:2007gi,Vanderveld:2008vi,Valkenburg:2009iw,Clifton:2009nv,Szybka:2010ky,2011JCAP...02..025B,Flanagan:2012yv,2013JCAP...12..051L,2015arXiv150706590L} or Szekeres~\cite{2009GReGr..41.1737B,2010PhRvD..82j3510B,2014PhRvD..90l3536P,2014JCAP...03..040T} holes.

When considered at first order, cosmological perturbations do not generate any bias in the cosmological observables, such as the distance-redshift relation~\cite{Valageas:1999ch}, by construction. However, they introduce a dispersion: some objects are more or less receding due to their peculiar velocities~\cite{2006PhRvD..73l3526H}, which brings a positive or negative correction to their cosmological redshift; some lines of sight are focused or defocused, which magnifies of demagnifies the images thus reducing or increasing the apparent distance to their sources. At second order in perturbations, things are much subtler and more technically involved~\cite{2005PhRvD..71f3537B}: one has now to take into account post-Born corrections, coupling between first-order perturbations, as well as genuinely second-order perturbations. The full second-order correction to the distance-redshift relation has been calculated independently by two groups. On the one hand, Ben-Dayan et al.~\cite{2012JCAP...11..045B,2013JCAP...11..019F,2015JCAP...08..020F} used a specific coordinate system, namely the geodesic-light-cone coordinates~\cite{2011JCAP...07..008G,2016JCAP...06..008F}, in order to facilitate the analysis of light propagation and light-cone averages. On the other hand, Umeh et al. performed the calculation directly from the standard coordinates of the cosmological perturbation theory~\cite{2014CQGra..31t2001U,2014CQGra..31t5001U}.

The latter results were used in ref.~\cite{2014JCAP...11..036C} to calculate the impact of second-order perturbations on the measurement of the distance to the last-scattering surface. The unexpected amplitude of this correction (percent level) seemed at odds with the old Weinberg conjecture~\cite{1976ApJ...208L...1W}, according to which the \emph{average} effect of gravitational lensing on the distance-redshift relation should vanish at any order, due to flux conservation. Although Weinberg's conjecture is not strictly correct~\cite{Ellis:1998ha,2013JCAP...12..051L,2016JCAP...06..008F}, a weaker version of this conjecture had been formulated later by Kibble \& Lieu~\cite{2005ApJ...632..718K}, who emphasised the crucial importance of (i) the choice of the observable that is averaged (distance, magnitude, luminous intensity, \ldots); and (ii) the notion of average that is used (angular average, area average, \ldots). This conclusion has been recently re-understood and generalised independently by Bonvin et al.~\cite{2015JCAP...06..050B,2015JCAP...07..040B} and Kaiser \& Peacock~\cite{2016MNRAS.455.4518K}.

However, refs.~\cite{2016MNRAS.455.4518K,2015JCAP...07..040B,2015JCAP...06..050B} and the additional works of Ben-Dayan et al.~\cite{2012JCAP...04..036B,2013JCAP...06..002B,2013PhRvL.110b1301B,2013arXiv1309.6542N} did not comprehensively address the issue of the second-order bias of the distance-redshift relation, in particular when the Hubble diagram is at stake. On the one hand, refs.~\cite{2016MNRAS.455.4518K,2015JCAP...07..040B,2015JCAP...06..050B} focused on the effect of gravitational lensing only, i.e. the deflection of light by matter overdensities, and thus did not consider the perturbations of the observed redshift, notably due to peculiar velocities. On the other hand, none of refs.~\cite{2016MNRAS.455.4518K,2015JCAP...07..040B,2015JCAP...06..050B,2012JCAP...04..036B,2013JCAP...06..002B,2013PhRvL.110b1301B,2013arXiv1309.6542N} investigated which notion of average is adapted to the observation of the Hubble diagram, which may differ from the most common angular or ensemble averages. The aim of the present article is to fill this gap, by fully evaluating the effect of second-order perturbations on the Hubble diagram, whether it is constructed from SN, QSO, or GW observations, and quantifying the corresponding impact on the inference of the cosmological parameters.

The outline is the following. In sec.~\ref{sec:averages_Hubble_diagram}, after having briefly explained why the choice of the right averaging procedure is crucial in order to accurately compare observations with theoretical models, we carefully derive the notion of average---source averaging---adapted to the observation of the Hubble diagram. In sec.~\ref{sec:bias_distance_observables} we calculate the bias of various source-averaged distance observables, i.e. the bias of the observed Hubble diagram for different choices of the distance indicator, at second order in cosmological perturbations. Finally, in sec.~\ref{sec:cosmological_parameters} we quantify the associated bias of the inferred cosmological parameters, and discuss how to reduce it. Note that, apart from sec.~\ref{sec:bias_distance_observables}, the article is intended to be accessible to non-theoretical cosmologists.

We use units in which the speed of light~$c$ is unity. Averaging operators are represented by angle brackets~$\ev{\cdots}$, with a subscript indicating the nature of the average. For instance $\ev{\cdots}_\Omega$ denotes angular average, $\ev{\cdots}_A$ area average, and the no-subscript~$\ev{\cdots}$ denotes ensemble average. They are all precisely defined in the text. The notation $\delta_x\define (x-\bar{x})/\bar{x}$ denotes the relative correction to a quantity~$x$ with respect to its value~$\bar{x}$ in a FLRW Universe; while $\Delta_x\define \ev{\delta_x}$ denotes a bias. A lower-case~$d$ denotes a distance, like angular distance~$d\e{A}$ or luminosity distance~$d\e{L}$; we use the upper-case~$D$ as a general notation for a distance indicator, like distance itself, but also magnitude~$m$ or luminous intensity~$I$.

\section{Averages and the Hubble diagram}
\label{sec:averages_Hubble_diagram}

\subsection{Observations, averaging, and lensing}

Most cosmological observations consists in, or at least rely on, a measurement of the mean relation between the redshift~$z$ and the distance~$D$ of luminous objects in our Universe~\cite{2015arXiv151103702F}. This is clearly the case for the Hubble diagram, whatever the nature of the source it is constructed from\footnote{This includes the sources of gravitational waves, which behave identically to light as far as those observables are concerned}; but also the anisotropies of the cosmic microwave background (CMB) radiation, where a key observable is the angular size~$\theta_*$ of the sound horizon at last scattering; as well as the imprint of the baryon acoustic oscillations (BAO) in the distribution of galaxies; the gas fraction of galaxy clusters; weak and strong lensing; and time delays. 

\begin{figure}[h!]
\centering
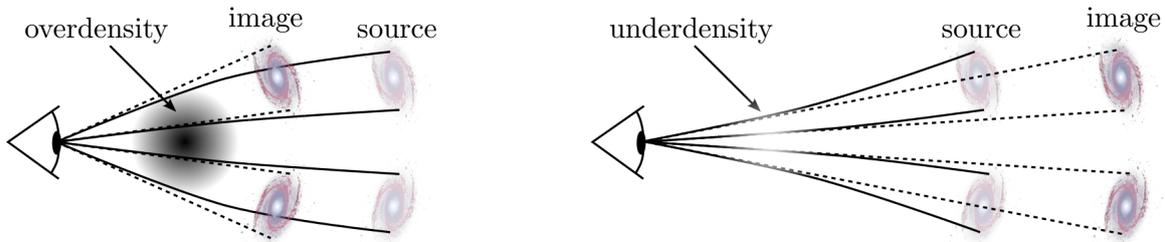
\caption{Effect of lensing on the apparent distance and angular separation of objects. Left panel: an overdensity provokes light focusing, i.e. magnification, making images seem bigger or closer to the observer, and further apart from one another. Right panel: on the contrary, an underdensity creates defocusing, i.e. demagnification, making images seem smaller or farther from the observer, but closer to one another.}
\label{fig:lensing}
\end{figure}

Roughly speaking, the observational strategy consists in collecting many data points $(z,D)$, directly or indirectly, and comparing them to the prediction of the FLRW model. This model indeed provides a formula $D(z|\{\Omega\})$ that can be used to fit the data and measure the cosmological parameters~$\{\Omega\}$. This procedure relies on the hypothesis that, even though our Universe is not perfectly homogeneous, \emph{on average} observations should match the homogeneous model. Albeit intuitive, this hypothesis raises subtle issues that we may want to account for, e.g., gravitational lensing. Gravitational lensing has two different observational consequences. On the one hand the global bending of light beams generates a remapping of the apparent positions of light sources on the sky. On the other hand, as light beams themselves are focused and sheared, the apparent morphology and luminosity of those light sources is affected. Of course those two phenomena are not independent, as they both stem from the same inhomogeneity of the matter distribution in the Universe. The practical result is that focused lines of sight get apparently pushed away with respect to each other, while defocused lines of sight get closer to each other (see fig.~\ref{fig:lensing}).

An important consequence of lensing is that the various observational notions of average that one can define are not equivalent. The canonical example is the difference between angular average and area average. For an observable quantity $Q$ (which can be angular distance, luminosity, number count, etc.), the angular average is
\begin{equation}
\ev{Q(z)}_\Omega  \define \frac{1}{4\pi} \int\e{sky}  Q(z,\vect{\theta}) \; \dd\Omega,
\end{equation}
and the area average is
\begin{equation}
\ev{Q(z)}_A\define \frac{1}{A} \int\e{sky} Q(z,\vect{\theta}) \; \dd A,
\end{equation}
where $A$ is the total area of an iso-$z$ surface centred on the observer. From the computational point of view, it is easy to relate those two notions of average by recalling that, by definition of the angular diameter distance~$d\e{A}$, one has $\dd A = d\e{A}^2 \dd\Omega$, therefore
\begin{equation}
\ev{Q}_A = \frac{\int\e{sky} Q(z,\vect{\theta}) d\e{A}^2(z,\vect{\theta})\;\dd\Omega}{\int\e{sky} d\e{A}^2(z,\vect{\theta}) \; \dd\Omega} 
					= \frac{\ev{Q\,d\e{A}^2}_\Omega}{\ev{d\e{A}^2}_\Omega}.
\end{equation}
From the intuitive point of view, this difference comes from the fact that, in angular average one gives the same weight to each direction in the observer's sky, whereas in area average one gives the same weight to each patch of the source surface. These two procedures do not agree as lensing precisely introduces differences of the apparent angular size $\Omega$ of two patches of the source surface that have the same physical area $A$, depending on whether they are magnified ($\Omega$ increases) or demagnified ($\Omega$ decreases).

\begin{figure}[h!]
\centering
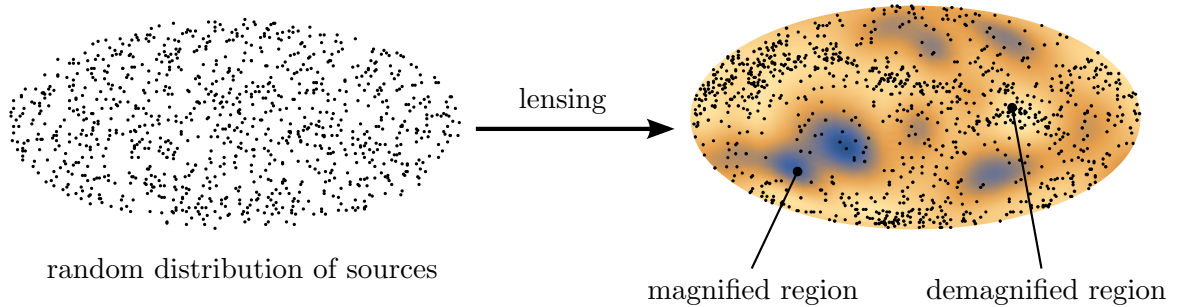
\caption{On the left, a random distribution of sources on the sky. On the right, the same distribution is lensed according to the superimposed magnification map. Blue regions indicate magnified regions, where images appear larger, brighter, and less dense on the sky, as if one had zoomed in. Light yellow regions indicate demagnified regions, where on the contrary images appear smaller, fainter, and with a denser distribution, as if one had zoomed out.}
\label{fig:remapping}
\end{figure}

This example illustrates the fact that one has to care about which notion of average is used to interpret the outcome of a survey. Suppose, in particular, that one is observing SNIa in order to measure the distance-redshift relation. To which average should we compare the data? Is it $\ev{d\e{A}}_\Omega(z)$, as advocated by ref.~\cite{2015JCAP...07..040B}? Or $\ev{d\e{A}}_A(z)$ as assumed in refs.~\cite{2012JCAP...04..036B,2013JCAP...06..002B,2013arXiv1309.6542N}? Or something else? We notice that it cannot be angular average for the following reason. Suppose for simplicity that SNIa are homogeneously distributed in space. Then, in absence of lensing, their distribution would look like the left panel of fig.~\ref{fig:remapping}. However, because of lensing, the actual distribution of their images on the observer's sky looks like the right panel of fig.~\ref{fig:remapping}: images tend to flee magnified regions and concentrate in demagnified regions. As a consequence, if one measures the distance-redshift relation by averaging over the observed SNIa, one gives more weight to the demagnified regions of the sky than to the magnified regions. The resulting average can thus not be a directional average, which gives each region of the sky the same weight. We will see that area average is not appropriate either.

More generally, we expect each cosmological observable, and each data analysis procedure, to require its own notion of average. In the remainder of this article we will focus on the case of the standard analysis of the Hubble diagram.

\subsection{The Hubble diagram: average over sources}

Consider a Hubble diagram, that is a set of observational data points~$\{(z_i,D_i)\}$, where $D$ is a distance indicator. In SN survey one usually uses the magnitude~$m$ (or distance modulus); in the GW case the luminosity distance~$d\e{L}$ seems to be preferred. This set of data points is fitted with the FLRW expression~$\bar{D}(z|\{\Omega\})$, depending parameters~$\{\Omega\}$ by minimizing
\begin{equation}
\chi^2(\{\Omega\}) \define \sum_{i=1}^N \pac{\frac{D_i - \bar{D}(z_i|\{\Omega\})}{\sigma_i}}^2,
\end{equation}
where $\sigma_i$ is the assumed uncertainty on $\{(z_i,D_i)\}$. Note that in reality the data analysis takes into account the correlation between different data points, due to the way data is collected and reduced, with a non-diagonal covariance matrix~$\mat{C}$. The $\chi^2$ is then defined as $\transpose{\mat{X}}\mat{C}^{-1}\mat{X}$, where $\vect{X}$ is a vector containing the data. We neglect this subtlety here.

If the theoretical model is adapted, then the best-fit expression~$D_*(z)\define \bar{D}(z|\{\Omega_*\})$ should reproduce the whole distance-redshift relation without favouring any particular redshift domain. In other words, it should match the binned data
\begin{equation}\label{eq:best_fit_average}
D_*(z) = \ev{D(z)}_N,
\end{equation}
where $\ev{D(z)}_N$ is obtained by averaging over the data points in the bin $\mathcal{B}_z$ containing $z$,
\begin{equation}\label{eq:average_redshift_bin}
\ev{D(z)}_N \define \frac{1}{N_z}\sum_{i\in\mathcal{B}_z} D_i,
\end{equation}
assuming that errors~$\sigma_i$ are all similar within a same redshift bin. This notion of average is know as \emph{source average}\footnote{One should not be confused about this expression, which does not mean that we give the same weight to each line of sight corresponding to some angular coordinates of the sources. Source averaging rather gives the same weight to the lines of sight in which sources are actually observed. Thus, from this point of view, it would be more appropriate to talk about \emph{image average}. However, the dichotomy between source and image is characteristic to a mindset associated with the perturbation theory, in which the respective positions of sources and images have a meaning, and can be distinguished. There is no such a dichotomy in general; we therefore use the expression source average in this article, keeping in mind that it means that each source (equivalently the associated image) has the same weight in the averaging procedure.}~\cite{2015MNRAS.454..280K}, or \emph{number-count average}. Let us now see how to calculate it from a theoretical model.

Suppose the catalogue contains a very large number of sources with a good sky coverage. Within the redshift bin~$\mathcal{B}_z$, we then have a large number $N_z$ of sources with a given distribution over the observed sky. Let us divide this sky into~$N_{\Omega}$ patches of equal size $\Delta\Omega$. Suppose that $\Delta\Omega$ is small enough so that we can consider the distance-redshift relation~$D(\vect{\theta},z)$ constant across each patch. The sum of eq.~\eqref{eq:average_redshift_bin} can then be reorganised as a sum over the directions~$\vect{\theta}_k$ of those sky patches as
\begin{equation}
\ev{D(z)}_N \define \frac{1}{N_z} \sum_{k=1}^{N_\Omega} N(z,\vect{\theta}_k)\,D(z,\vect{\theta}_k),
\end{equation}
where $N(\vect{\theta}_k,z)$ denotes the number of SNIa in $\mathcal{B}_z$ observed in the sky patch around $\vect{\theta}_k$. Multiplying this equation by $\Delta\Omega/\Delta\Omega$, and taking the limit $\Delta\Omega\rightarrow 0$, we can turn this discrete sum into an integral as
\begin{equation}\label{eq:sum_to_integral}
\ev{D(z)}_N
= \sum_{k=1}^{N_{\Omega}}
\frac{N(z,\vect{\theta}_k)}{N_z \Delta\Omega}\,D(z,\vect{\theta}_k)\, 
\Delta\Omega
\rightarrow
\int\e{sky} p(z,\vect{\theta}) 
D(z,\vect{\theta}) \; \dd\Omega
\end{equation}
with the probability density function~$p(z,\vect{\theta})$, such that $p(z,\vect{\theta})\dd\Omega$ represents the probability of observing an image of the bin~$\mathcal{B}_z$ within the infinitesimal sky patch of size $\dd\Omega$ around $\vect{\theta}$. Equivalently, $N_z\,p(z,\vect{\theta})\dd\Omega$ is the number of images within this patch. In the following, it will turn out to be more convenient to work with the number density~$n(z,\vect{\theta})\define N_z p(z,\vect{\theta})$, in terms of which eq.~\eqref{eq:sum_to_integral} reads
\begin{equation}
\ev{D(z)}_N
=
\frac{\int\e{sky}D(z,\vect{\theta}) \, n(z,\vect{\theta}) \; \dd\Omega}
{\int\e{sky} n(z,\vect{\theta}) \; \dd\Omega}
=
\frac{\ev{D n}_{\Omega}}{\ev{n}_\Omega}.
\end{equation}
Note that the denominator is nothing but $N_z$.

\begin{figure}[h!]
\centering
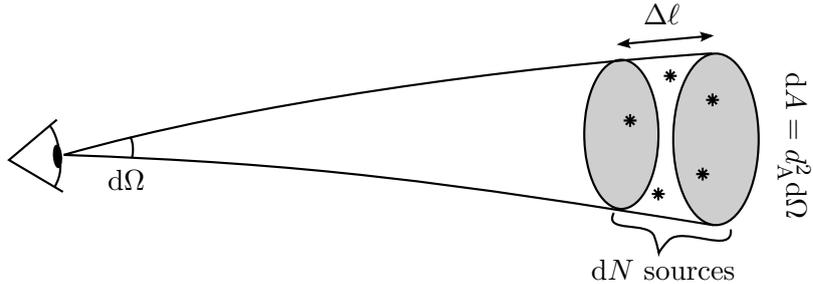
\caption{Element of volume~$\dd V=\Delta\ell\,\dd A$ and sources corresponding to an observed patch of the sky of size $\dd\Omega$ in a redshift bin~$\Delta z$.}
\label{fig:sky_patch}
\end{figure}

Let us now evaluate $n(z,\vect{\theta})$. As illustrated in fig.~\ref{fig:sky_patch}, the infinitesimal interval mentioned above corresponds to a physical volume~$\dd V = \dd A\times \Delta\ell$. On the one hand, $\dd A = d\e{A}^2 \dd\Omega$ by definition of the angular-diameter distance $d\e{A}$. On the other hand, the physical depth~$\Delta\ell$ of the redshift bin reads
\begin{equation}
\Delta\ell=\frac{\Delta z}{(1+z) H_{||}},
\end{equation}
where $H_{||} \define (k^\mu k^\nu \nabla_{\mu} u_\nu)/(1+z)^2$ is the local longitudinal expansion of the cosmological fluid, $u^\mu$ being the local four-velocity of matter in the Universe and $k^\mu$ the wave four-vector of light.

\newcommand{\area}{\dd A = d\e{A}^2 \dd\Omega}

In the volume element~$\dd V$, there are $\dd N = \rho\e{s} \dd V$ sources, where $\rho\e{s}$ is the volume density of sources as measured in their rest frame. Replacing $\dd V$ by its expression, in terms of angles and redshift, we get
\begin{equation}
\dd N = \frac{\rho\e{s} d\e{A}^2 \Delta z}{(1+z) H_{||}} \,\dd\Omega,
\end{equation}
and since this number is also~$\dd N = n(z,\vect{\theta})\dd\Omega$, we finally get the theoretical expression of the source average~$\ev{\cdots}_N$ as
\begin{empheq}[box=\fbox]{equation}\label{eq:source_average_theoretical}
\ev{D(z)}_N
=
\frac{\int\e{sky} D(z,\vect{\theta}) \, \rho\e{s} d\e{A}^2 H_{||}^{-1} \; \dd\Omega}
{\int\e{sky} \rho\e{s} d\e{A}^2 H_{||}^{-1} \; \dd\Omega} .
\end{empheq}
This formula is valid in any spacetime. We stress again that this notion of average is only adapted to the interpretation of the Hubble diagram. It is not suited for other observables such as the CMB or BAO. In particular, since the analysis of the CMB is based on angular power spectra~$C_\ell$, it is rather sensitive to directional averages. The conclusions of ref.~\cite{2015JCAP...06..050B} regarding the effect of second-order lensing on the CMB are therefore independent from the current work.

Although it not very explicit in eq.~\eqref{eq:source_average_theoretical}, it is understood that the fields~$\rho\e{s}$ and $H_{||}$ in the integrand are evaluated \emph{at the end of the light rays} (i.e. null geodesic) corresponding to the observation directions~$\vect{\theta}$ across a surface of constant redshift~$z$. In other words, in order to evaluate the integrals, one must in principle be able to solve the equations of geometric optics in our cosmological model in order to determine the map $(z,\vect{\theta})\mapsto x\e{s}^\mu$, where $x\e{s}^\mu$ are the coordinates of the source event.

\subsection{Physical interpretation and limitations}

\begin{figure}[h!]
\centering
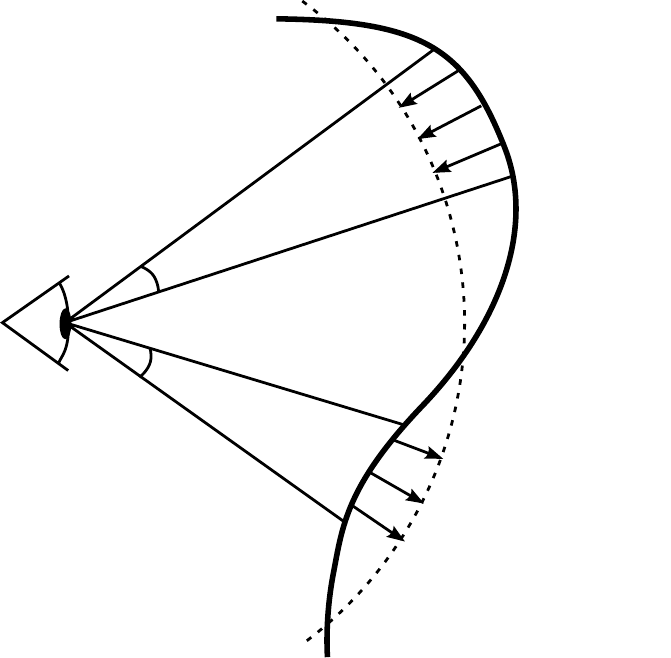
\hspace{1cm}
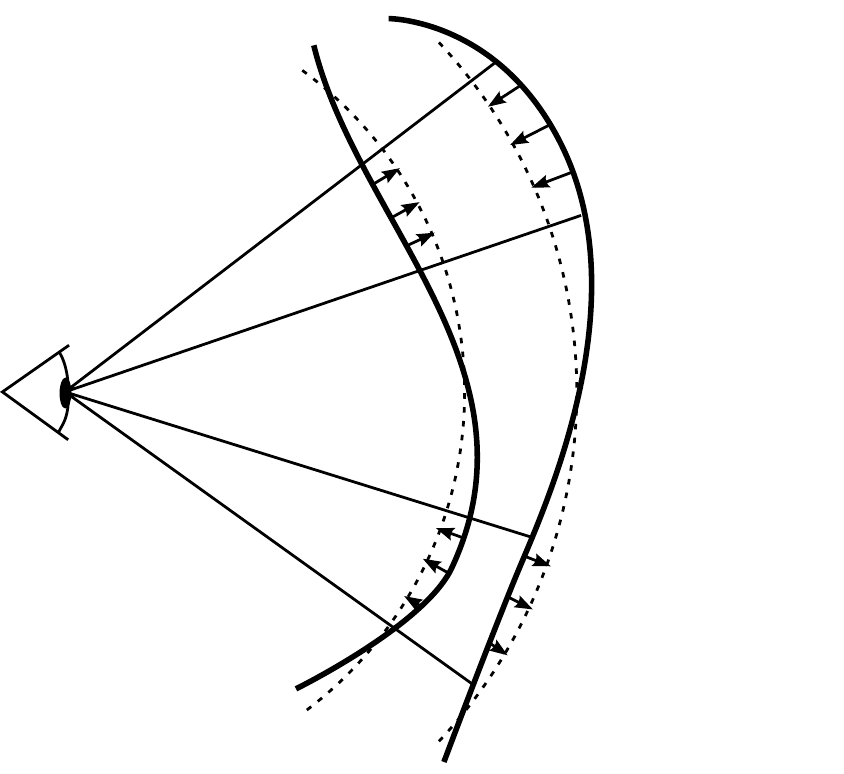\hspace*{-1.5cm}
\caption{Effect of peculiar velocities on the relation between redshift space and physical space. \emph{Left panel}: effect of peculiar velocities on the geometry of an iso-$z$ surface in physical space (thick solid line). The dotted line corresponds to an iso-$D$ surface with respect to the observer, or equivalently to the iso-$z$ surface in a FLRW model. \emph{Right panel}: idem but for two iso-$z$ surfaces, illustrating the relation between the local longitudinal expansion rate~$H_{||}$ and the physical thickness~$\Delta\ell$ of a redshift bin~$\Delta z$.}
\label{fig:peculiar_velocities}
\end{figure}

Before we jump to the calculation of~$\ev{D}_N$ at second order in cosmological perturbations, let us discuss the physical meaning of the integration kernel~$\rho\e{s}d\e{A}^2 H_{||}^{-1}$ of eq.~\eqref{eq:source_average_theoretical}. All those terms are indeed associated to well-known physical effects in cosmology.
\begin{description}
\item[Spatial distribution of the sources.] The term~$\rho\e{s}$ allows for the intrinsic inhomogeneity of the spatial distribution of the sources, regardless of light propagation. It naturally gives more weight, in the average of over sources~$\ev{\cdots}_N$, to regions of the Universe where sources are more abundant.
\item[Gravitational lensing.] As already discussed above, $d\e{A}^2$ connects the apparent size~$\Omega$ of a patch of the sky to the corresponding physical area~$A$ over which sources are distributed. Along a line of sight that is, e.g., demagnified---i.e. if light mostly goes through underdense regions of the Universe---a given $\Omega$ corresponds to larger-than-average $A$, where potentially more sources are present, hence enhancing the weight of this line of sight. The converse applies for magnified lines of sight.
\item[Peculiar velocities.] Because $d\e{A}$ is actually $d\e{A}(z)$, any effect that influences redshifts, mostly peculiar velocities but also Sachs-Wolfe (SW) and integrated Sachs-Wolfe (ISW) or Rees-Sciama effects, are accounted for. The effect of peculiar velocities is easy to interpret (see fig.~\ref{fig:peculiar_velocities}): consider two regions~$R_1$, $R_2$, measured with the same redshift $z$ and under the same solid angle~$\Omega$, but such that $R_1$ has a peculiar velocity towards the observer while $R_2$ has the opposite. Because peculiar velocities add to the Hubble flow, this means that $R_1$ is actually farther than $R_2$, $d\e{A}(R_1)>d\e{A}(R_2)$, and thus corresponds to a larger area $A_1>A_2$, where potentially more sources are present. This phenomenon is sometimes called ``Doppler magnification'', and its cross-correlation with galaxy number counts has recently been suggested as a novel cosmological observable~\cite{2014MNRAS.443.1900B,2016arXiv161005946B}.

\item[Redshift-space distortions.] The physical phenomena affecting redshifts have also an impact, via their longitudinal gradient, on the conversion between the redshift width of a bin and the corresponding physical depth. This is illustrated in fig.~\ref{fig:peculiar_velocities} in the case of peculiar velocities: a faster longitudinal expansion~$H_{||}$ implies a larger ratio between the redshift separation~$\Delta z$ and distance separation~$\Delta\ell$ of sources. Note that all the other corrections to the redshift (SW, ISW, \ldots) are also accounted for by $H_{||}$, because this local expansion is defined with respect to the rest frame of the sources, a coordinate system for which any effect on the redshift is seen as a velocity.
\end{description}

Finally, it is important emphasise the assumed conditions under which the expression~\eqref{eq:source_average_theoretical} for $\ev{D(z)}_N$ to match the best-fit~$D_*(z)$ of an experimental Hubble diagram.
\begin{description}
\item[Unbiased uncertainties.] In eq.~\eqref{eq:average_redshift_bin} we did not take into account the fact that each data point $(z_i,D_i)$ is weighted by the inverse square of its uncertainty~$1/\sigma_i^2$. This is justified if $\sigma$ is a function of $z$ only, which is not necessarily the case. We however still recover eq.~\eqref{eq:average_redshift_bin} if the uncertainties are unbiased, i.e. if they do not favour deviations of $D$ towards particular a particular direction. Astrophysical phenomena are likely to produce such a bias.
\item[Comprehensive and balanced sky coverage.] The fact that $\ev{\cdots}_N$ involves an integral over the whole sky implicitly assumes that the survey to which it is compared also covers the whole sky, instead of focusing on a given region as it is currently the case for most SN surveys~\cite{2014A&A...568A..22B}. In the latter situation, one has to multiply the integration kernel by $(\dd\tau\e{obs}/\dd\Omega)(\vect{\theta})$, representing the fraction of observation time spent in the direction~$\vect{\theta}$.
\item[Large number of sources.] The transition~\eqref{eq:sum_to_integral} from a discrete sum to an integral requires to have a large number of sources, so that the sky can be safely divided into small enough patches where the distance-redshift can be considered mostly homogeneous. The impact of increasing the size of such patches can be observed in fig.~5 of ref.~\cite{2016arXiv160804403G}.
\end{description}

\section{Bias of distance observables}
\label{sec:bias_distance_observables}

This section is dedicated to the computation of the bias of source-averaged distance indicators~$\ev{D}_N$ at second order in cosmological perturbations. We remind the reader that $D$ can stand for angular or luminosity distance, luminous intensity (sometimes called flux), or magnitude. What we call \emph{bias} here is the difference between $\ev{D(z)}_N$, i.e. the observed average distance-redshift relation, and the standard~$\bar{D}(z)$ predicted by an unperturbed FLRW model with the same cosmological parameters. In other words, such a bias quantifies the error that one makes when fitting the Hubble diagram with the standard relation~$\bar{D}(z)$.

In what follows, we will always consider the fractional bias defined as
\begin{equation}\label{eq:bias_indicator_definition}
\Delta_D \define \frac{\ev{D}_N-\bar{D}}{\bar{D}} = \ev{ \frac{D-\bar{D}}{\bar{D}} }_N \define \ev{\delta_D}_N ,
\end{equation}
except when $D$ represents the magnitude~$m$, which is already a logarithmic quantity, so that we rather use $\delta_m \define m-\bar{m}$ and $\Delta_m \define \ev{\delta_m}_N$ in that case. Furthermore, because we do not know what is the actual spacetime geometry of the Universe but only its statistical properties, the only quantity that we can predict is the \emph{ensemble average}~$\ev{\Delta_D}$ of $\Delta_D$, i.e. its average over multiple realisation of the Universe. Note that, as extensively discussed in ref.~\cite{2015JCAP...07..040B}, it is crucial to take this ensemble average \emph{after} source-averaging, since they do not commute in general.

\subsection{Calculation at second order}
\label{subsec:second_order_bias}

\subsubsection{Preliminaries}

We aim at using the results of ref.~\cite{2014CQGra..31t2001U,2014CQGra..31t5001U}, which provided a comprehensive analysis of light propagation up to second order in cosmological perturbations, and determined the expression of the angular distance-redshift relation~$d\e{A}(z)$ at that order. The first step therefore consists in expressing $\Delta_D$ in terms of this result. We thus consider~$D$ a function of the angular distance~$d\e{A}$, so that, expanding $D(\bar{d}\e{A}+\delta d\e{A})$ at second order yields
\begin{equation}\label{eq:delta_D_and_Delta_D_A}
\delta_D = \alpha \delta_{d\e{A}} + \beta (\delta_{d\e{A}})^2
\end{equation}
with
\begin{equation}
\alpha \define \frac{\bar{d}\e{A}}{\bar{D}} \ddf{D}{d\e{A}}
\qquad \text{and} \qquad
\beta \define \frac{1}{2} \frac{\bar{d}\e{A}^2}{\bar{D}} \ddf[2]{D}{d\e{A}},
\end{equation}
except, again, when $D=m$ in which case eq.~\eqref{eq:delta_D_and_Delta_D_A} still holds but with slightly different definitions for $\alpha$ and $\beta$ (remove $\bar{D}$). Table~\ref{tab:distance_indicators} lists the values of $\alpha$ and $\beta$ associated with the most commonly used distance indicators.

\begin{table}[t]
\centering
\begin{tabular}{l|cc}
distance indicator~$D$ & $\alpha$ & $\beta$ \\ 
\hline 
angular, luminosity distances $d\e{A},d\e{L}$ & 1 & 0 \\ 
luminous intensity~$I\propto d\e{L}^{-2}$ & -2 & 3 \\ 
magnitude~$m=5\log_{10} d\e{L}+\cst$ & $\frac{5}{\ln 10}$ & $\frac{-5}{2\ln 10}$ \\ 
\end{tabular}
\caption{Parameters of the expansion~\eqref{eq:delta_D_and_Delta_D_A} for various distance indicators~$D$.}
\label{tab:distance_indicators}
\end{table}

Similarly to the distance indicator~$D$, the averaging kernel~$n\propto \rho\e{s} d\e{A}^2 H_{||}^{-1}$ can also be expanded perturbatively as $n=\bar{n}(1+\delta_n)$, so that
\begin{equation}
\Delta_D = \ev{\delta_D}_N 
= \frac{\ev{n \delta_D}_\Omega}{\ev{n}_\Omega} 
= \frac{\ev{(1+\delta_n) \delta_D}_\Omega}{\ev{1+\delta_n}_\Omega}
= \ev{\delta_D}_\Omega + \ev{\delta_D \delta_n}_\Omega + \order(3),
\end{equation}
where we neglected the apparently order-two term~$\ev{\delta_D}_\Omega\ev{\delta_n}_\Omega$ which is, actually, of order four because first-order perturbations vanish once averaged over the sky. Since $\delta_n$ is only involved multiplied by $\delta_D$, we only need its expression at first order:
\begin{equation}
\delta_n = b\e{s}\delta + 2\delta_{d\e{A}} - \delta_{H_{||}}.
\end{equation}
In the above expression, we assumed that the density contrast~$\delta\e{s}$ of the sources (SNIa, quasars, GW) is proportional to the total-matter density contrast~$\delta$~\cite{2016arXiv161109787D}, with a bias $b\e{s}$ that must be estimated from astrophysics or simulations. In the remainder of this article we will only keep the dominant contributions to~$\delta_n$. Its complete expression at first order can be found in refs.~\cite{2011PhRvD..84f3505B,2011PhRvD..84d3516C}.

Putting everything together, the bias of the distance indicator~$D$ measured from the Hubble diagram reads
\begin{equation}\label{eq:bias_indicator_perturbative}
\Delta_D = 
\alpha \ev{\delta_{d\e{A}}}_\Omega 
+ (2\alpha + \beta) \ev{(\delta_{d\e{A}})^2}_\Omega
+ \alpha b\e{s} \ev{\delta_{d\e{A}}\delta}_\Omega
- \alpha\ev[2]{\delta_{d\e{A}}\delta_{H_{||}}}_\Omega.
\end{equation}
Except for the first term, all the $\delta_{\cdots}$ corrections can be evaluated at first order.

\subsubsection{Cross-terms}

Let us start by evaluating the quadratic terms, for which only first-order perturbations are required. In the following we will focus on the dominant contributions only, i.e. gravitational lensing and peculiar velocities, leaving the so-called GR corrections, or horizon-scale corrections (gravitational redshifts, ISW,\ldots) aside.

In this context the first-order correction to the angular distance is the sum of the lensing convergence~$\kappa$ and the so-called velocity convergence~$\kappa_v$ (described on the left of fig.~\ref{fig:peculiar_velocities}),
\begin{equation}
\delta_{d\e{A}} = -\kappa_v - \kappa
\end{equation}
with
\begin{equation}
\kappa_v \define \pa{ \frac{1}{\Hc \chi} - 1 } v_\chi
\qquad \text{and} \qquad
\kappa \define 4\pi G\bar{\rho}_0 \int_0^\chi \frac{\chi'(\chi-\chi')}{\chi} \frac{\delta(\eta',\chi'\vect{\theta})}{a(\eta')}
\end{equation}
where $\Hc$ is the conformal Hubble rate, $\chi$ the comoving coordinate distance between the source and the observer, $v_\chi$ the radial component of the source's peculiar velocity, and in the expression of $\kappa$, $\eta'=\eta_0-\chi'$ is the conformal time when the photon is at a comoving distance~$\chi'$ far from the observer. Besides, the leading fractional correction to the local longitudinal expansion is simply
\begin{equation}
\delta_{H_{||}} = \frac{\partial_\chi v_\chi}{\mathcal{H}}.
\end{equation}

Among the various cross-terms that appear in the expression~\eqref{eq:bias_indicator_perturbative}, two vanish by symmetry:
\begin{equation}
\ev{\kappa_v\delta}_\Omega = \ev{\kappa_v \delta_{H_{||}}}_\Omega = 0,
\end{equation}
the orientation of the peculiar velocity (inwards or outwards) is indeed uncorrelated with the density contrast or the expansion when those quantities are evaluated at the exact same location\footnote{This does not hold when the quantities are evaluated at different locations~$\vect{x}$, $\vect{y}$. In this case the correlation function displays a characteristic dipolar structure which can be used as an independent cosmological probe~\cite{2014MNRAS.443.1900B,2016arXiv161005946B}.}. The other terms, once ensemble-averaged, can be written as
\begin{equation}
\ev{X}_\Omega = \int \frac{\dd k}{H_0} \, \kernel_{X}(\eta,k) \, \mathcal{P}_\delta(\eta,k),
\end{equation}
where $\mathcal{P}_\delta(\eta,k)$ denotes the dimensionless power spectrum of density fluctuations at conformal time~$\eta$, i.e. such that
$\ev{ \delta(\eta,\vect{k}) \delta(\eta,\vect{k}') } 
= (2\pi)^3 \delta\e{D}(\vect{k}-\vect{k}') 2\pi^2 k^{-3} \mathcal{P}_\delta(\eta,k)$,
while~$\kernel_{X}$ is a dimensionless integration kernel, whose expression for each cross-term is
\begin{align}
\kernel_{\kappa_v^2} 
&= \frac{H_0 f^2}{3k^3} \pa{\frac{1}{\chi}-\Hc}^2 \\
\kernel_{\kappa_v\kappa} ,
&= \frac{3f}{2k^2}\pa{\frac{1}{\chi}-\Hc}\frac{H_0^3 \Omega\e{m0}}{1+z}
\int_0^\chi \dd \chi' \; \frac{g(\eta')}{g(\eta)} \, \frac{\chi'(\chi-\chi')}{\chi} \, j_1[k(\chi-\chi')], \\
\kernel_{\kappa \delta} &= \frac{3}{2 k} H_0^3\Omega\e{m0} (1+z)
\int_0^\chi \dd\chi' \; \frac{g(\eta')}{g(\eta)}\,\frac{\chi'(\chi-\chi')}{\chi}\, j_0[k(\chi-\chi')],\\
\kernel_{\kappa \delta_{H_{||}}} &= \frac{3f}{2k} H_0^3 \Omega\e{m0} (1+z) \int_0^\chi \dd\chi' \;
\frac{g(\eta')}{g(\eta)} \, \frac{\chi'(\chi-\chi')}{\chi} \, j_2[k(\chi-\chi')],\\
\kernel_{\kappa^2} &= \pac{ \frac{3}{2} H_0^2 \Omega\e{m0} (1+z) }^2 \frac{H_0}{k} \int_0^\chi \dd\chi_1 \int_0^\chi \dd\chi_2 \; \frac{\chi_1\chi_2(\chi-\chi_1)(\chi-\chi_2)}{\chi^2} \nonumber\\
&\hspace{8cm}\times \frac{g(\eta_1) g(\eta_2)}{g^2(\eta)} \, j_0 [k(\chi_1-\chi_2)] .
\end{align}
In the above equations, $g$ is the linear growth function of the gravitational potential; $f\define \dd\ln (a g)/\dd\ln a \approx \Omega\e{m}^{0.55}(z)$ is the linear growth rate; and $j_n(x)\define (-1)^n \dd^n (\sin x/x)/\dd x^n$ denotes the spherical Bessel function of order~$n$.

\subsubsection{Sky-averaged angular distance}

We now turn to the first term~$\ev{\delta_{d\e{A}}}_\Omega$ in eq.~\eqref{eq:bias_indicator_perturbative}. As shown in ref.~\cite{2014JCAP...11..036C}, based on the full second-order calculations of ref.~\cite{2014CQGra..31t5001U}, the leading correction to the \emph{ensemble average} angular distance reads
\begin{equation}\label{eq:ensemble_average_distance}
\ev{\delta_{d\e{A}}}
=
\Delta\e{loc} + \Delta\e{int},
\end{equation}
that is the sum of a local term---due to peculiar velocities and gravitational potential---and an integrated term---due to gravitational lensing,
\begin{align}
\Delta\e{loc} &\define \bigg\langle \pa{ \frac{\Hc'}{2\Hc^2} - \frac{1}{\Hc\chi} -\frac{1}{2} } 
															v_\chi^2 + \chi \partial_\chi\Phi \, v_\chi	
													+ \pa{ 1 - \frac{1}{\Hc\chi} }
														\pa{ \chi v_\chi v_\chi' + v_{\perp}^2 + \chi \Phi\,\partial_\chi v_\chi }  \bigg\rangle,\\
\Delta\e{int} &\define \frac{3}{2} \ev{\kappa^2} .
\end{align}

Now, as emphasised by ref.~\cite{2015JCAP...07..040B}, ensemble average~$\ev{\cdots}$ is not quite the same thing as directional average~$\ev{\cdots}_\Omega$. The actual meaning of ensemble averaging in refs.~\cite{2014CQGra..31t2001U,2014CQGra..31t5001U,2014JCAP...11..036C,2015JCAP...07..040B} is very subtle, and we refer the reader to appendix~\ref{app:ensemble_average} for a detailed discussion. For the present purpose the only important element is that the ensemble average of an observable quantity~$Q$, in the sense of refs.~\cite{2014CQGra..31t2001U,2014CQGra..31t5001U,2014JCAP...11..036C,2015JCAP...07..040B}, is related to its directional average as
\begin{equation}\label{eq:relation_angular_ensemble_averages}
\ev{Q} = \ev{Q}_\Omega - 2\ev{\kappa Q}_\Omega.
\end{equation}
The above equation is valid up to second order in perturbation theory. Note that because $\kappa$ is of order 1 in cosmological perturbations, if $Q$ is a correction of order $n$, then the difference~$\ev{Q}-\ev{Q}_\Omega$ is at least of order $n+1$. Since we work at order 2, this means that we do not have to worry about the distinction between ensemble and directional averages for quantities that are already second order; for example we can consider that $\ev{\kappa^2}=\ev{\kappa^2}_\Omega$. \emph{Therefore, in the following we will always use the simple notation~$\ev{\cdots}$ when averaging second-order quantities}.

Back to the correction to the angular distance, by combining eqs.~\eqref{eq:ensemble_average_distance}, \eqref{eq:relation_angular_ensemble_averages} we get
\begin{equation}
\ev{\delta_{d\e{A}}}_\Omega = \Delta\e{loc} - 2\ev{\kappa \kappa_v} - \frac{1}{2} \ev{\kappa^2}.
\end{equation}

\subsubsection{Results}

\begin{figure}[t]
\centering
\includegraphics[width=\textwidth]{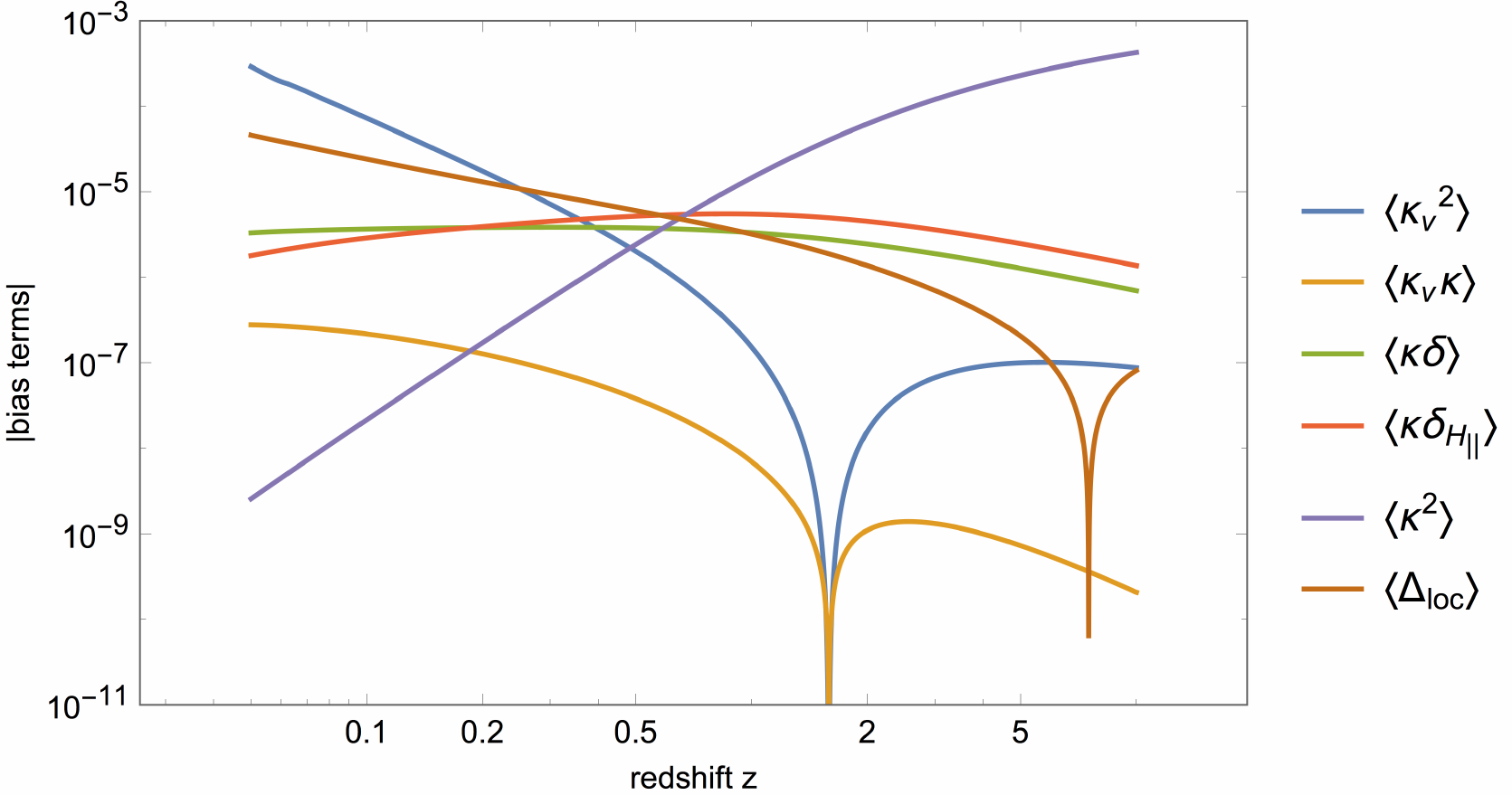}
\caption{Comparison of the various terms involved in the bias~$\Delta_D$ of the distance indicator~$D$ measured from the Hubble diagram.}
\label{fig:comparison_bias_terms}
\end{figure}

We have seen that the bias of the distance indicator~$\Delta_D$ contains six different terms: $\ev{\kappa_v^2}, \ev{\kappa_v\kappa}, \ev{\kappa\delta}, \ev[2]{\kappa\delta_{H_{||}}}, \ev{\kappa^2}, \ev{\Delta\e{loc}}$. We compare in fig.~\ref{fig:comparison_bias_terms} the amplitude of those terms and their dependence on redshift. Following ref.~\cite{2014JCAP...11..036C}, since we are interested in times that are much later the matter-radiation equality, we parametrised the matter power spectrum as
\begin{equation}
\mathcal{P}_\delta(z,k) = A \pa{ \frac{k}{k_0} }^{n\e{s}-1} \pac{\frac{2}{5} \, \frac{G(z)}{G_\infty} \frac{k^2 T(k)}{H_0^2 \Omega\e{m0}}}^2
\end{equation}
where $A=2.2\times 10^{-9}$ is the amplitude of primordial curvature perturbation at the pivot scale~$k_0=0.05\U{Mpc}^{-1}$; $n\e{s}=0.96$ is the spectral index characterizing the degree of scale-invariance of primordial perturbations; $G_\infty=G_0(3+2\Omega\e{m0}^{-0.45})/5$; and $T(k)$ denotes the transfer function, which describes the effect of the radiation-matter transition on the amplitude of the various perturbation modes. We used the following fitting function for the growth factor~\cite{PeterUzan},
\begin{equation}
G(z) = \frac{5}{2(1+z)} \,
\frac{\Omega\e{m}(z)}{\Omega\e{m}^{4/7}(z)-\Omega_\Lambda(z)+[1-\Omega\e{m}(z)/2][1+\Omega_\Lambda(z)/70]},
\end{equation}
and we relied on the fitting formula proposed by ref.~\cite{1998ApJ...496..605E} to compute $T(k)$. Finally, we used the best-fit $\Lambda$CMB \textsl{Planck} cosmological parameters~\cite{2016A&A...594A..13P}: $H_0=100 h\U{km/s/Mpc}$ with $h=0.67$, $\Omega\e{b0}h^2=0.022$, $\Omega\e{c0}h^2=0.12$, and $\Theta_{2.7}=1.0094$~\cite{2009ApJ...707..916F}.

As expected, on the one hand the local terms associated with peculiar velocities, such as $\ev{\kappa_v^2}$ and $\Delta\e{loc}$, dominate at low redshift due to their prefactors containing~$1/\chi$. On the other hand, the pure-lensing term~$\ev{\kappa^2}$ dominates at high redshift, with a transition around~$z=0.5$. This behaviour is in agreement with earlier works~\cite{2013PhRvL.110b1301B}.

We also note that $\ev{\kappa_v\kappa}$ is always smaller by 3 orders of magnitude than the dominant term at any redshift. We will therefore neglect it in the following, and write
\begin{empheq}[box=\fbox]{equation}\label{eq:bias_distance_indicator_result}
\Delta_D = \alpha \pa{ \Delta\e{loc} - b\e{s} \ev{\kappa\delta} + \ev[2]{\kappa \delta_{H_{||}}} } 
					+ (2\alpha+\beta) \ev{\kappa_v^2} 
					+ \pa{\frac{3\alpha}{2} + \beta} \ev{\kappa^2}.
\end{empheq}

%

\subsection{Comparing various distance indicators}

The fact that $\Delta_D$ in eq.~\eqref{eq:bias_distance_indicator_result} depends explicitly on~$\alpha$, $\beta$ shows that the Hubble diagram is differently biased depending on the distance indicator that is fitted. This dependence is due to purely statistical considerations: of course each observation~$(z_i,D_i)$ is affected the same way by inhomogeneities, regardless of the choice of $D$; it is when all those observations are put together and averaged over that the choice of $D$ becomes critical. 

The lensing contribution~$\ev{\kappa^2}$ to $\Delta_D$, which dominates at high redshifts, vanishes when $3\alpha/2+\beta=0$, i.e. if $D\propto d\e{A}^{-2}\propto I$. This property was first noticed in ref.~\cite{2005ApJ...632..718K} and recently confirmed in refs.~\cite{2013JCAP...06..002B,2013PhRvL.110b1301B,2015JCAP...07..040B,2016MNRAS.455.4518K,2015JCAP...08..020F} in a more general context. At low redshift the dominant contribution~$\ev{\kappa_v^2}\propto 1/z^2$, vanishes when $2\alpha+\beta=0$, i.e. for $D\propto d\e{A}^{-3}$ as found by ref.~\cite{2015MNRAS.454..280K}. For intermediate redshifts all contributions are comparable and there is not simple expression of $D$ for which the associated bias cancels. However, this is not a serious issue as it remains very small ($\sim 10^{-5}$).

\begin{figure}[h!]
\centering
\includegraphics[width=0.8\textwidth]{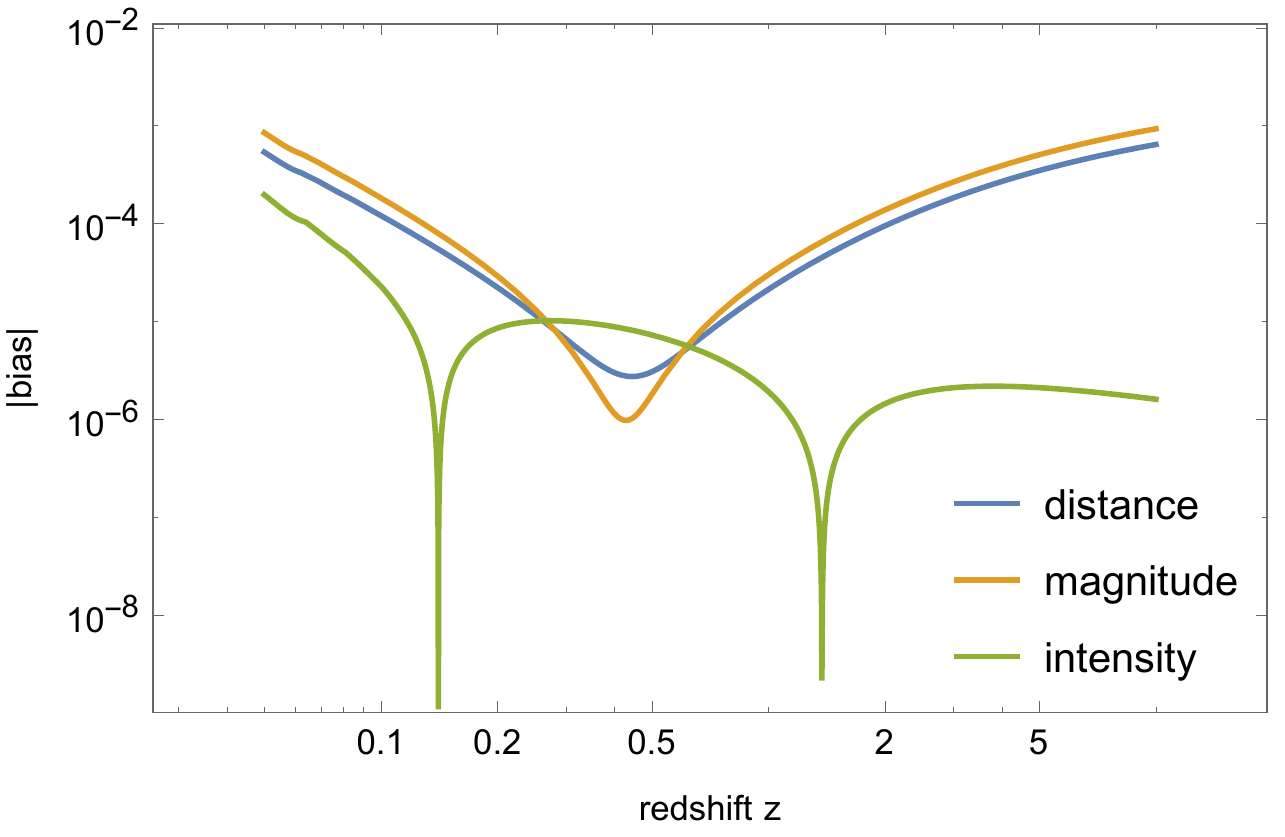}
\caption{Comparison of the (absolute value of the) bias~$\Delta_D$ of the Hubble diagram constructed with the distance indicator~$D$, for three different choices of $D$: distance (luminosity or angular); magnitude; and intensity. Both $\Delta_{d\e{A}}$ and $\Delta_{m}$ are positive, while $\Delta_I$ is first negative, then positive, and negative again.}
\label{fig:comparison_bias_indicators}
\end{figure}

In fig.~\ref{fig:comparison_bias_indicators} we compare the bias of the most common distance indicators, namely angular or luminosity distance, magnitude, and luminous intensity. We set $b\e{s}=1$ for simplicity, but other choices do not significantly change the results. Among the three choices for~$D$, intensity~$I\sim d\e{L}^{-2}$ is clearly the least biased one, except at intermediate redshifts ($z\sim 0.5$). The bias~$\Delta_D$ remains smaller than~$10^{-3}$ for any choice, but $\Delta_I\leq 10^{-5}$ for $z\geq 0.5$. In that redshift domain, one can therefore reduce the bias of the Hubble diagram due to inhomogeneities by a factor 100 simply by fitting~$I(z)$ instead of $m(z)$.

\section{Impact on the inference of cosmological parameters}
\label{sec:cosmological_parameters}

Even if the bias of the Hubble diagram due to second-order perturbations remains quite small, it can have an impact on the inference of the cosmological parameters that is significant enough from the point of view of precision cosmology, especially for high-redshift surveys.

\subsection{Effective dark-energy equation of state}
\label{subsec:weff}

The dark-energy equation of state~$w$, relating the effective pressure~$p\e{DE}$ and energy density~$\rho\e{DE}$ of dark energy, modelled as a perfect fluid, via $p\e{DE}=w \rho\e{DE}$ has a rather weak impact on the distance-redshift relation. In a FLRW Universe with zero spatial curvature ($K=0$), we have
\begin{equation}
d\e{A}(z) = \frac{1}{(1+z)} \int_0^z \frac{\dd \zeta}{H(\zeta)},
\end{equation}
and if the dark-energy equation of state is a general function of redshift~$w(z)$, Friedmann's equations impose
\begin{equation}\label{eq:Hubble_rate_DE}
\frac{H^2(z)}{H_0^2} = \Omega\e{m0} (1+z)^3 + (1-\Omega\e{m0}) \exp\pac{ 3 \int_0^z \frac{1+w(\zeta)}{1+\zeta} \; \dd\zeta },
\end{equation}
where we see that the detailed evolution of $w(z)$ is not crucial, as what is involved is essentially a form of redshift average of it. This implies that, on the contrary, any attempt to reconstruct~$w(z)$ from the observation of $d\e{A}(z)$ is very sensitive to its possible biases. Differentiating eq.~\eqref{eq:Hubble_rate_DE} one finds~\cite{2010PhRvL.104u1301C}
\begin{equation}\label{eq:w_z}
1+w(z) = \frac{1+z}{3} \, \frac{3\Omega\e{m0}(1+z)^2 d'^3+2 d''}{\Omega\e{m0}(1+z)^3 d'^3-d'},
\end{equation}
with $d(z)\define H_0 (1+z) d\e{A}(z)$ and where a prime denotes a derivative with respect to $z$.

\begin{figure}[h!]
\centering
\includegraphics[width=0.49\textwidth]{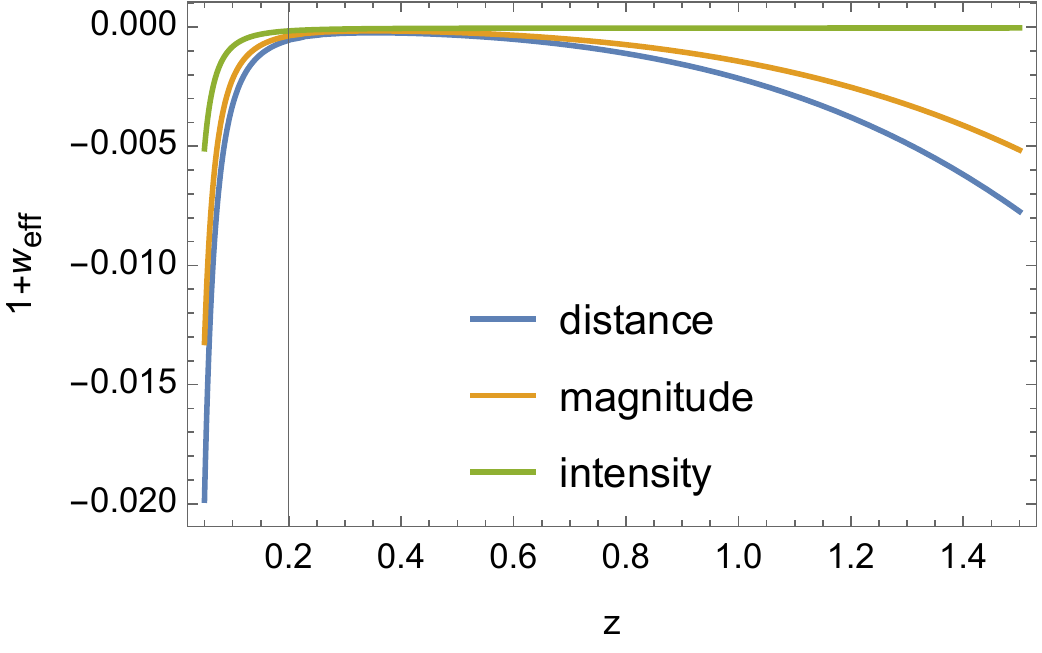}
\hfill
\includegraphics[width=0.49\textwidth]{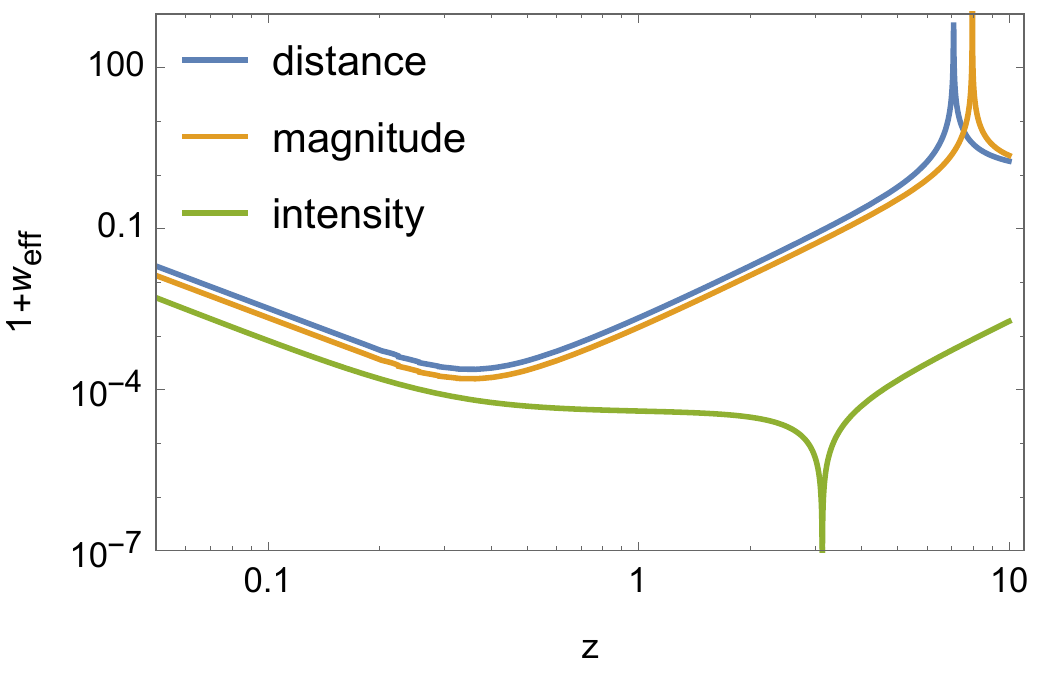}
\caption{Effective dark-energy equation of state~$w\e{eff}(z)$ obtained in a $\Lambda$CDM Universe ($w=-1$) caused by the bias~$\Delta_D$ on the distance indicator~$D$ fitted from the Hubble diagram. The left panel corresponds to the typical redshift range of SN surveys.}
\label{fig:weff}
\end{figure}

Now if $d(z)$ is the result of a fit of the Hubble diagram, or calculated from another distance indicator which is itself a fit of the Hubble diagram, then the observational bias~$\Delta_D$ will generate a spurious correction to~$w(z)$. In fig.~\ref{fig:weff} we plot the effective dark-energy equation of state~$w\e{eff}(z)$ calculated from eq.~\eqref{eq:w_z} with a $d(z)$ reconstructed from three different biased indicators~$D$, namely
\begin{align}
\text{with distance:} \qquad d(z) &= \bar{d}(z)\pac{1+\Delta_{d\e{A}}(z)}, \\
\text{with magnitude:} \qquad d(z) &= \bar{d}(z)\, 10^{\frac{\Delta_m(z)}{5}}, \\
\text{with intensity:} \qquad d(z) &= \bar{d}(z) \sqrt{\frac{1}{1+\Delta_I(z)}},
\end{align}
where $\bar{d}(z)$ is the standard FLRW expression of $d(z)$ in a $\Lambda$CDM model i.e. with $w=-1$.

As expected, $w\e{eff}(z)$ appears as a quantity which is particularly sensitive to the corrections of the distance-redshift relation. Even for the relatively small redshift range corresponding to SN surveys ($z\leq 1.5$), the corrections are of percent level. When higher redshift are probed, the difference between $w\e{eff}$ and the true~$w$ is beyond unity if the distance indicator is distance or magnitude; it remains sub-percent for intensity. This shows that it is crucial to choose a weakly biased distance indicator in Hubble diagrams reaching high redshifts, such as those constructed with QSOs or standard sirens, if one wants to reconstruct the history of the dark-energy equation of state.

\subsection{Biased cosmological parameters}

\begin{figure}[h!]
\centering
\includegraphics[width=0.7\textwidth]{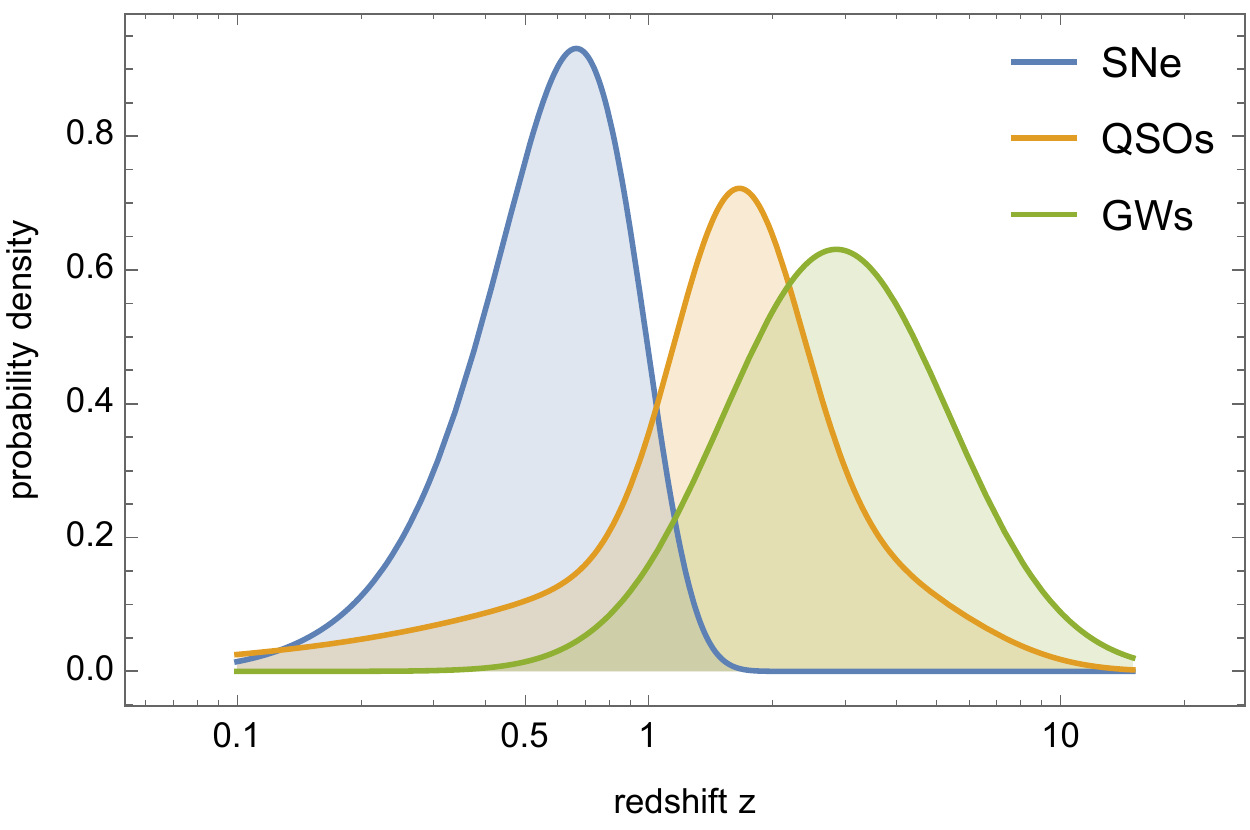}
\caption{Logarithmic redshift distribution~$p_{\log}(z)=z p(z)$ of sources for the three different kinds of surveys considered here: an LSST-like SN survey, a quasar (QSO) survey, and a LISA-like gravitational wave (GW) survey.}
\label{fig:redshift_distributions}
\end{figure}

Let us now turn to an estimation of the bias of the cosmological parameters inferred from the Hubble diagram. We consider three types of surveys with very different redshift ranges:
\begin{enumerate}
\item a LSST-like~\cite{2009arXiv0912.0201L} supernova survey (SNIa) with $z<1.2$;
\item a quasar survey (QSOs) inspired from ref.~\cite{2015ApJ...815...33R} with $z<6$;
\item an eLISA-like~\cite{2016JCAP...10..006C} standard siren survey (GWs) with $z<10$.
\end{enumerate}
The corresponding redshift distributions $p(z)$ are compared in fig.~\ref{fig:redshift_distributions}.

In order to estimate the bias on cosmological parameters for each of those surveys, we consider mock binned data~$(z_i,D_i)$, where $D_i$ is simply given by
\begin{equation}
D_i \define \bar{D}(z_i)\times[1+\Delta_D(z_i)],
\end{equation}
the barred quantity~$\bar{D}$ being evaluated for a fiducial background $\Lambda$CDM model with the same parameters~$\{\bar{\Omega}\}$ as in sec.~\ref{subsec:second_order_bias}. For SNIa we use 12 redshift bins of width $\Delta z=0.1$ from $0$ to $1.2$; for QSOs 21 bins from $0.05$ to $7$ in log scale; and for GWs 9 bins of width~$\Delta z=1$ from $z=0$ to $z=9$. For each bin $\mathcal{B}_i=[z_{\mathrm{min},i};z_{\mathrm{max},i}]$ we define $z_i$ as the barycentre of $\mathcal{B}_i$ weighted by $p(z)$,
\begin{equation}
z_i \define \frac{1}{z_{\mathrm{max},i}-z_{\mathrm{min},i}} \int_{z_{\mathrm{min},i}}^{z_{\mathrm{max},i}} z\, p(z) \; \dd z.
\end{equation}

We then fit the resulting mock Hubble diagrams~$(z_i,D_i)$ by minimizing
\begin{equation}
\chi^2_D (\{\Omega\})
\propto
\sum_{i=1}^{N\e{bin}}
p(z_i) \pac{ D\e{mod}(z_i|\{\Omega\}) - D_i }^2
\end{equation}
with two different models. On the one hand we consider a $w_0 w_a$CDM model, where the dark-energy equation of state~$w$ is allowed to vary with redshift according to
\begin{equation}\label{eq:w0_wa}
w = w_0 + w_a (1-a) = w_0 + \frac{w_a \, z}{1+z},
\end{equation}
$w_0, w_a$ being two constants. On the other hand, we consider dark energy as a cosmological constant ($w=-1$) but allow spatial curvature~$K$ to be non vanishing ($\Lambda K$CDM). The resulting best-fit cosmological parameters~$\Omega^*$ are summarised in table~\ref{tab:best_fit_parameters}, for the three usual choices for $D$: luminosity distance~$d\e{L}$, magnitude~$m$, and intensity~$I$.

\begin{table}[h!]
\centering
\begin{tabular}{|c|c|ccc|cc|}
\cline{3-7} 
\multicolumn{2}{c|}{} & \multicolumn{3}{c|}{$w_0 w_a$CDM} & \multicolumn{2}{c|}{$\Lambda K$CDM} \\ 
\hline 
survey & $D$ &$\Omega\e{m0}^*-\bar\Omega\e{m0}$ & $w_0^*+1$ & $w_a^*$ 
& $\Omega\e{m0}^*-\bar\Omega\e{m0}$ & $\Omega_{K0}^*$ \\ 
\hline
\rowcolor{lightgray}
\cellcolor{white}
\multirow{3}{*}{SNIa}
& $d\e{L}$ 
& $-4.5\times 10^{-4}$ & $4.3\times 10^{-4}$ & $4.6\times 10^{-3}$ & $-2.7\times 10^{-4}$ & $4.7\times 10^{-4}$ \\ 
& $m$ 
& $-6.1\times 10^{-4}$ & $5.4\times 10^{-4}$ & $7.0\times 10^{-3}$ & $4.9\times 10^{-5}$ & $-1.2\times 10^{-4}$\\ 
& $I$ 
& $-2.0\times 10^{-3}$ & $2.6\times 10^{-3}$ & $1.7\times 10^{-2}$ & $1.0\times 10^{-3}$ & $-1.7\times 10^{-3}$ \\ 
\hline 
\multirow{3}{*}{} 
& $d\e{L}$
& $-9.0\times 10^{-4}$ & $-9.7\times 10^{-4}$ & $2.0\times 10^{-2}$ & $-1.1\times 10^{-3}$ & $2.4\times 10^{-3}$ \\ 
\rowcolor{lightgray}\cellcolor{white}QSOs
& $m$
& $-5.2\times 10^{-4}$ & $4.0\times 10^{-5}$ & $8.4\times 10^{-3}$ & $-5.3\times 10^{-4}$ & $1.0\times 10^{-3}$ \\ 
& $I$
& $-6.0\times 10^{-3}$ & $8.0\times 10^{-3}$  & $5.1\times 10^{-2}$ & $5.9\times 10^{-3}$ & $-9.3\times 10^{-3}$ \\ 
\hline
\multirow{3}{*}{GWs} 
& $d\e{L}$
& $-1.0\times 10^{-3}$ & $-2.7\times 10^{-3}$ & $3.0\times 10^{-2}$ & $-1.4\times 10^{-3}$ & $3.1\times 10^{-3}$ \\ 
& $m$
& $-6.0\times 10^{-4}$ & $-4.6\times 10^{-4}$ & $1.2\times 10^{-2}$ & $-7.2\times 10^{-4}$ & $1.5\times 10^{-3}$\\ 
\rowcolor{lightgray}\cellcolor{white}
& $I$
& $-3.3\times 10^{-6}$ & $3.7\times 10^{-5}$ & $-1.2\times 10^{-4}$ & $-2.2\times 10^{-5}$ & $4.8\times 10^{-5}$\\ 
\hline 
\end{tabular} 
\caption{Cosmological parameters~$\Omega^*$ fitted from a biased Hubble diagram~$D(z)$, depending on the distance indicator~$D$ (luminosity distance~$d\e{L}$, magnitude~$m$, or intensity~$I$), for three different kinds of survey (SNIa, QSOs, GWs), and with two different cosmological models ($w_0w_a$CDM or $\Lambda K$CDM). We highlighted the least biased distance indicator for each survey.}
\label{tab:best_fit_parameters}
\end{table}

We can see that the bias on cosmological parameters is always relatively small: at most at percent level for $w_a$, and always sub-percent for the other parameters. This is at least one order of magnitude lower than the uncertainties on the cosmological parameters expected from LSST~\cite{2009arXiv0912.0201L}, eLISA~\cite{2016JCAP...10..006C}, or obtained by the current QSO Hubble diagram~\cite{2015ApJ...815...33R}.

Such a result could seem in contradiction with subsec.~\ref{subsec:weff}, where we found that the inferred dark-energy equation of state can have order-unity departures from $w=-1$ at high redshift. This apparent discrepancy can be understood by recalling that fitting the observed $D(z)$ is not equivalent to fitting $w(z)$ directly. In particular, $D(z)$ depends on an integral of $w(z)$, which makes it sensitive to its whole behaviour from $0$ to $z$. As a consequence, it is impossible to intuitively guess the best-fit values of $w_0$, $w_a$ by fitting $w\e{eff}(z)$ with $w_0+w_a z/(1+z)$. We can see from fig.~\ref{fig:weff} that $w\e{eff}(z)$ is very close to $-1$ between $z=0.1$ and $z=0.8$. Somehow $D(z)$ knows about it, even at high redshift. This tends to prevent any fit of $D(z)$ using $(w_0, w_a)$ to deviate too much from $1+w_0=w_a=0$. From these results we conclude that model-independent studies, using principal component analysis for the estimation of $\{\Omega\e{m0},w(z)\}$ like in refs.~\cite{2005PhRvD..71b3506H,2014PhRvD..90f3006N,2017APh....86....1Z}, are thus expected to be \emph{much more biased}---at a level comparable to the $w\e{eff}(z)$ found in subsec.~\ref{subsec:weff}---than analyses using the standard $(w_0,w_a)$ parametrisation.

Finally, it is interesting to note that each survey has a different least biased distance indicator: luminosity distance for SNIa, magnitude for QSOs (although the difference between distance and magnitude is very small) and intensity for GWs. It is easy to understand this result by comparing~$\Delta_D(z)$ of fig.~\ref{fig:comparison_bias_indicators} with the survey's redshift distributions~$p(z)$ of fig.~\ref{fig:redshift_distributions}. The distributions of both SNIa and QSOs peak around $z=0.5-1$, which is contained in the small window where $|\Delta_I| > |\Delta_{d\e{A}}|, |\Delta_{m}|$. On the contrary, the GW distribution peaks at much higher redshift, where $|\Delta_I|  < |\Delta_{d\e{A}}|, |\Delta_{m}|$ as it is free from gravitational lensing. It was therefore expected that the biases weighted by the redshift distributions would lead to such results.

\subsection{How to remove the bias in practice?}

Even though this bias of the Hubble diagram due to cosmological perturbations turns out to be relatively small, it is still a systematic effect that will eventually need to be corrected as the precision of observations increases. There are in practice two possibilities to get rid of the bias. The brute-force idea consists in calculating it for each distance indicator and systematically subtracting it from the observations. This may require to establish fitting formulae for~$\Delta_D$ for any set of cosmological parameters, in order to improve computing efficiency.

Another approach, approximate but completely free and less model-dependent, consists in choosing the right distance indicator to plot and fit the Hubble diagram at hand. As we can see from table~\ref{tab:best_fit_parameters}, in the case of GW choosing to fit~$I(z)$ instead of $d\e{L}(z)$ allows one to reduce the bias by two orders of magnitude, which is a significant gain. More clever choices of the quantities~$\alpha, \beta$ can further reduce the bias if needed. This method is also particularly efficient for direct reconstructions of the dark-energy equation of state~$w(z)$, where the error reaches unity for a traditional Hubble diagram~$m(z)$, but reduces to $10^{-3}$ when $I(z)$ is fitted instead.

\section{Conclusion}
\label{sec:conclusion}

In this article, we have investigated how the inhomogeneity of the Universe, modelled by the cosmological perturbation theory, biases the distance-redshift relation probed by the Hubble diagram. Our additions, compared to previous works on the same topic, are (i) the careful analysis of the type of averaging procedure---source averaging---that is hidden in the standard analysis of the Hubble diagram; and (ii) the full calculation of the associated bias of the average distance-redshift relation at second order in cosmological perturbations, for various distance indicators, including all the relevant effects, namely peculiar velocities, inhomogeneity in the source distribution, redshift-space distortions, and gravitational lensing.

Our results can be summarised as follows:
\begin{itemize}
\item The fractional bias~$\Delta_D$ on any of the usual distance indicators~$D$ remains smaller than $10^{-3}$ in the redshift range $0.05<z<10$; its exact value and evolution with redshift depends on the choice of $D$.
\item Although it is small, this bias is \emph{significant} for direct measurements of the evolution of the dark-energy equation of state~$w(z)$. Indeed, in a Universe with a cosmological constant ($w=-1$) we observe a spurious evolution of $w$, such that $1+w\e{eff}(z)\sim 1$ at high redshift. This can however be avoided by using luminous intensity~$I\propto d\e{L}^{-2}$ in the Hubble diagram instead of magnitude or distance.
\item The impact of $\Delta_D$ on the inference of the standard cosmological parameters remains very weak, in particular it is at least one order of magnitude lower than observational uncertainties. This also includes the dark-energy equation of state when the latter is parametrised according to~$w(a)=w_0+(1-a)w_a$. Summarising, the standard fit of the Hubble diagram is not significantly jeopardised by the cosmic inhomogeneity.
\end{itemize}

Note that besides the conditions stated at the very end of sec.~\ref{sec:averages_Hubble_diagram}, these conclusions are trustworthy to the extent that the inhomogeneity of the Universe can be modelled with the cosmological perturbation theory at second order. This excludes in principle the effect of very small scales, where the validity of the fluid limit itself should break down. This brings a range of additional effects~\cite{Futamase:1989hba,2011arXiv1109.2314C,2013PhRvD..87l3526F,2015JCAP...11..022F}, potentially non-negligible~\cite{2013PhRvL.111i1302F}, which must be further understood and tackled together with the effect of the large-scale structure investigated here. 


\section*{Acknowledgements}
We wish to thank Obinna Umeh for fruitful discussions throughout the evolution of this project. We acknowledge financial support from the National Research Foundation (NRF). Opinions expressed and conclusions arrived at, are those of the authors and are not necessarily to be attributed to the NRF. PF and RM are also supported by the South African SKA Project. RM is also supported by the UK STFC, Grant ST/N000668/1. 

\appendix

\section{Ensemble averaging}
\label{app:ensemble_average}

In its general sense, the ensemble average of a stochastic variable~$X$ is an average over a large number of realisations~$\{X_1,X_2,X_3,\ldots\}$, that is
\begin{equation}
\ev{X} \define \lim{N}{\infty} \frac{1}{N} \sum_{n=1}^\infty X_n.
\end{equation}
In the case of cosmology, what are considered stochastic are the small departures from homogeneity generated at the end of inflation. Ensemble averaging in cosmology thus corresponds to a though experiment in which we generate one Universe, perform one observation, then take another Universe, perform the same observation and so on. The subtlety of this thought experiment is that it requires to specify what remains fixed in the process. For example, in the case of the distance-redshift relation, how do you compare two sources in two different Universes? Do you consider two sources at the same redshift and observed in the same direction? Or rather two sources with the same background coordinates? Such a choice is comparable to a gauge issue, and it is entangled with the perturbative scheme at hand.

\begin{figure}[h!]
\centering
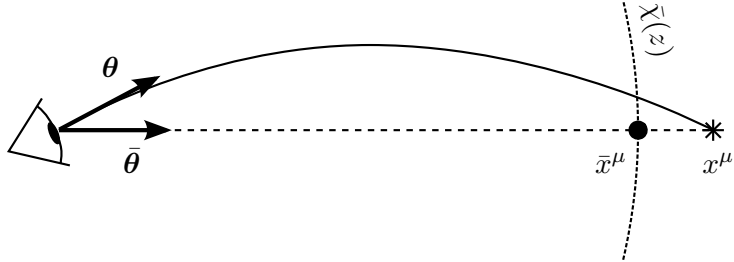
\caption{Definition of the quantities~$\vect{\theta}$, $\bar{\vect{\theta}}$, and $\bar{\chi}(z)$ involved in eq.~\eqref{eq:expansion_angular_distance}. A given redshift~$z$ and direction of observation~$\vect{\theta}$ are associated with a source event with coordinates~$x^\mu$ in the perturbed Universe, indicated by a star. The black disk indicates the coordinates of the associated background event~$\bar{x}^\mu$, which is maintained fixed in ensemble averages.}
\label{fig:ensemble_average}
\end{figure}

In refs.~\cite{2014CQGra..31t2001U,2014CQGra..31t5001U,2014JCAP...11..036C,2015JCAP...07..040B}, the angular distance-redshift relation is expanded as
\begin{equation}\label{eq:expansion_angular_distance}
d\e{A}(z,\vect{\theta}) 
= \frac{\bar{\chi}(z)}{1+z} \pac{ 1 + \delta^{(1)}_{d\e{A}}(\bar{\chi},\bar{\vect{\theta}}) 
															+ \delta^{(2)}_{d\e{A}}(\bar{\chi},\bar{\vect{\theta}}) 
															+ \order(3)},
\end{equation}
where $\bar{\chi}$ is the comoving distance to a source with redshift~$z$ in a FLRW Universe, $\vect{\theta}$ is the observed direction of the source, and $\bar{\vect{\theta}}$ are the background angular coordinates of the source, i.e. the direction in which it would be observed in a FLRW Universe. The meaning of those quantities is summarized in fig.~\ref{fig:ensemble_average}.

\begin{figure}[h!]
\centering
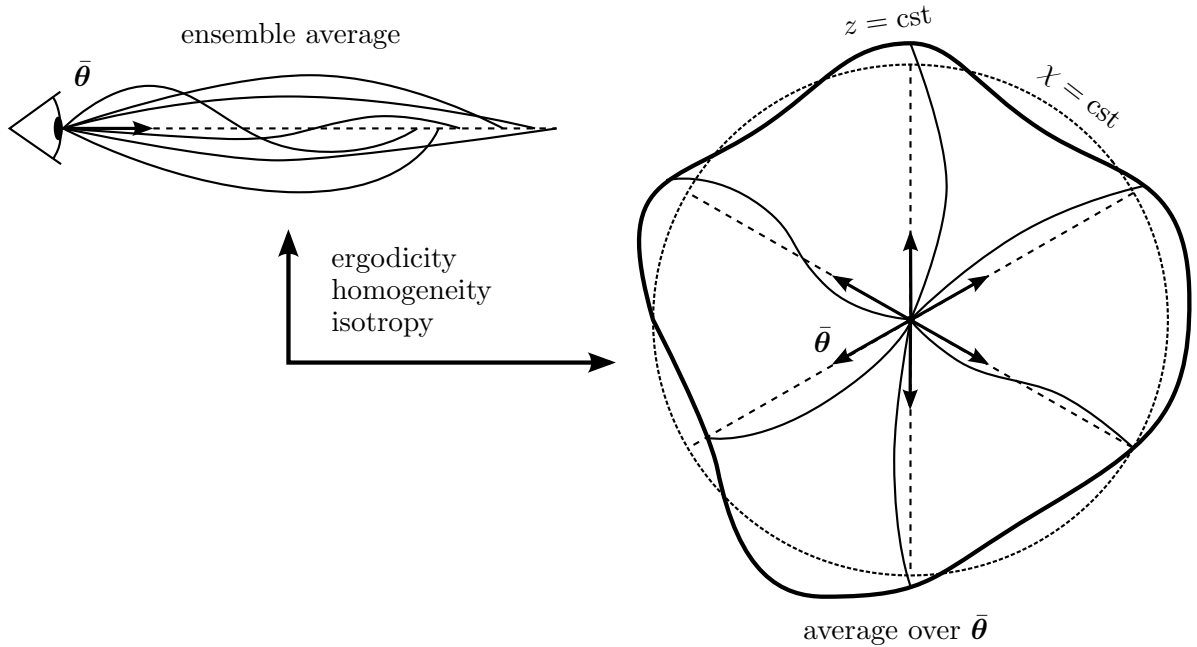
\caption{Correspondence between ensemble average and average over the background angular position~$\bar{\vect{\theta}}$ of sources.}
\label{fig:ergodicity}
\end{figure}

In these articles, when ensemble average is performed, what remains fixed is~$z$ (hence~$\bar{\chi}$) and $\bar{\vect{\theta}}$. The associated thought experiment thus looks like the left panel of fig.~\ref{fig:ergodicity}, which, assuming statistical homogeneity and isotropy of our Universe plus an ergodic principle, corresponds to the right panel of the same figure. Indeed, as long as only one-point distributions are concerned, any coordinate direction~$\bar{\vect{\theta}}$ can be considered a different realisation of the Universe. It is therefore clear that ensemble averaging corresponds to averaging over background directions across an iso-$z$ surface,
\begin{equation}
\ev{d\e{A}(z)} 
= \frac{1}{4\pi} \int\e{sky}  d\e{A}[z,\vect{\theta}(\bar{\vect{\theta}})] \; \dd\bar{\Omega}.
\end{equation}

This actually holds for any observable quantity~$Q$, and using~$\dd\bar\Omega =(1-2\kappa)\dd\Omega$, we find
\begin{equation}
\ev{Q}
= \frac{1}{4\pi} \int\e{sky} Q[z,\vect{\theta}(\bar{\vect{\theta}})] \; \dd\bar{\Omega}
= \frac{1}{4\pi} \int\e{sky} (1-2\kappa) \, Q(z,\vect{\theta})\; \dd\Omega
= \ev{Q}_\Omega - 2\ev{\kappa Q}_\Omega,
\end{equation}
which justifies eq.~\eqref{eq:relation_angular_ensemble_averages} in the main text. This is an alternative proof of the one proposed in ref.~\cite{2015JCAP...07..040B}, and also its generalisation to a general situation where perturbations of the affine parameter-redshift relation are allowed.

\bibliographystyle{JHEP.bst}
\bibliography{bibliography_Hubble_diagram.bib}


\end{document}